\documentclass[letterpaper,twocolumn,showpacs,preprintnumbers,amsmath,amssymb,floatfix,superscriptaddress,pre]{revtex4}

\usepackage{graphicx}  
\usepackage{dcolumn}   
\usepackage{bm}        
\usepackage{epsfig}
\usepackage{textcomp}
\usepackage[sort&compress]{natbib}

\begin{document}

\title{Distributed delays stabilize neural feedback systems}

\author{Ulrike Meyer}
\affiliation{%
Institute for Biology II, RWTH, 52074 Aachen, Germany}%

\author{Jing Shao}
\affiliation{
Department of Physics, Washington University in St.~Louis, MO 63130-4899, USA}

\author{Saurish Chakrabarty}
\affiliation{
Department of Physics, Washington University in St.~Louis, MO 63130-4899, USA}

\author{Sebastian F.~Brandt}
\affiliation{
Department of Physics, Washington University in St.~Louis, MO 63130-4899, USA}

\author{Harald Luksch}
\affiliation{Institute for Zoology, Technical University Munich, 85350 Freising-Weihenstephan, Germany}

\author{Ralf Wessel}
\email{rw@physics.wustl.edu}
\affiliation{
Department of Physics, Washington University in St.~Louis, MO 63130-4899, USA}

\date{November 30, 2007}
\begin{abstract}
We consider the effect of distributed delays in neural feedback systems. The avian optic tectum is reciprocally connected with the nucleus isthmi. Extracellular stimulation combined with intracellular recordings reveal a range of signal delays from 4 to 9 ms between isthmotectal elements. This observation together with prior mathematical analysis concerning the influence of a delay distribution on system dynamics raises the question whether a broad delay distribution can impact the dynamics of neural feedback loops. For a system of reciprocally connected model neurons, we found that distributed delays enhance system stability in the following sense. With increased distribution of delays, the system converges faster to a fixed point and converges slower toward a limit cycle. Further, the introduction of distributed delays leads to an increased range of the average delay value for which the system's equilibrium point is stable. The enhancement of stability with increasing delay distribution is caused by the introduction of smaller delays rather than the distribution per \nolinebreak[4] se.
\end{abstract}
\pacs{87.19.L-, 87.18.Sn, 87.10.-e}
\maketitle
%
\section{Introduction}
The signal flow in the brain is not just feedforward; rather, feedback dominates most neural pathways \cite{Shepherd}. Often pairs of reciprocally connected neurons are spatially separate by several millimeters. For instance, the primate corticothalamic feedback loop extends over a distance of approximately 100 mm. Thus, for a typical action potential speed of 1 mm/ms we expect a signal delay of 100 ms. When signal delays are larger than the neural response time, complex loop dynamics emerge \cite{Foss96,Foss97,Foss00}. 

For reciprocally connected populations of neurons, large delays can introduce another dimension, namely the distribution of delay times. Such a distribution could be an epiphenomenon in the evolution of larger brains, or it could be of adaptive significance. Work from applied mathematics states an influence of the distribution of delay times on system dynamics \cite{Gopalsamy,Bernard,Eurich02,Atay,Thiel,Eurich05}. Intrigued by the latter possibility, we asked two questions: What is the distribution of delay times in an experimentally accessible neural feedback system? What is the impact of distributed delays on a mathematically tractable neural model feedback system?

\begin{figure}[t]
  \begin{center}
    \epsfig{file=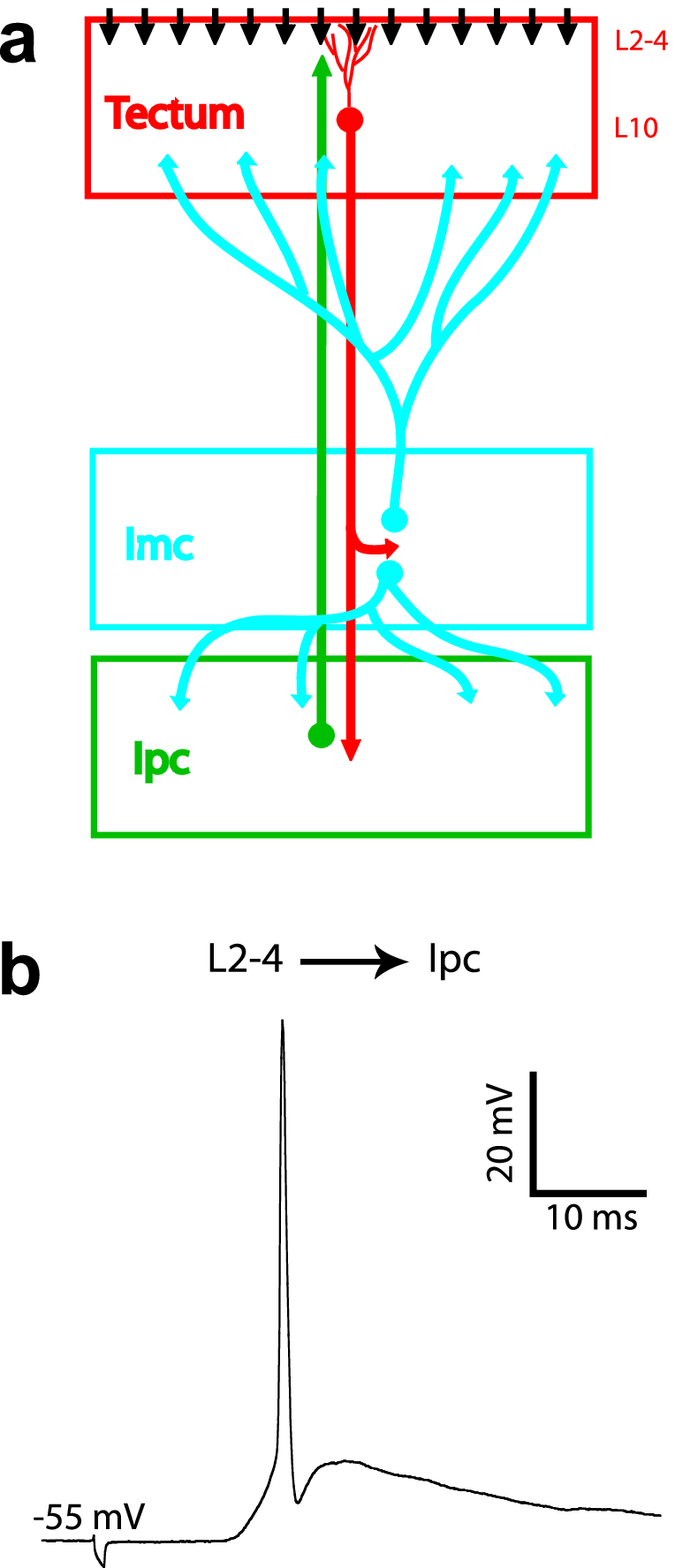,width=1.5in}    
    \caption{(Color) Schematic of the isthmotectal circuitry and representative response to electrical stimulation. {\bf (a)} Schematic of the isthmotectal circuitry. RGC axons (black arrows) enter in upper tectal layers. A subpopulation of tectal L10 neurons (red) projects to Imc and Ipc. The Imc nucleus consists of two populations of neurons (blue); one projecting broadly back to lower tectal layers and one projecting broadly to the Ipc nucleus. Ipc neurons (green) project back to the tectum with axons reaching into upper tectal layers. {\bf (b)} Intracellular recording from an Ipc neuron in response to electrical stimulation in tectal L2-4.}
    \label{fig1}
  \end{center}
\end{figure}

We measured the distribution of delay times in the isthmotectal feedback system of birds [Fig.~\ref{fig1}(a)] \cite{Luksch03,Wang03}. The avian isthmic nuclei (parabigeminal nucleus in mammals) receive a topographically organized projection from the tectum (superior colliculus in mammals), to which they project back and have been conjectured to mediate spatiotemporal attentional mechanisms \cite{Gruberg,Maczko,Marin07}. The isthmic nuclei in birds consist of three substructures: pars parvocellularis (Ipc), pars magnocellularis (Imc), and pars semilunaris (SLu) that are spatially separated from the tectum \cite{Wang04,Wang06}. In response to visual stimulation, the Ipc neurons undergo a transition from quiescence to rhythmic firing \cite{Marin07,Marin05}. Delays can drive a neural feedback loop over a stability boundary resulting in oscillatory behavior \cite{Marcus,Babcock,Laing,Brandt06b,Brandt07a,Brandt07b}. To elucidate the impact of a delay distribution on the system dynamics, we investigated, through numerical simulations and mathematical analysis, a model of reciprocally coupled neurons with distributed delays.

\section{Measured distribution of delays}
To measure the signal delays between pairs of isthmotectal elements, we obtained intracellular whole-cell recordings from identified neurons in a midbrain slice preparation and stimulated groups of presynaptic neurons or axons with brief electrical pulses delivered extracellularly [Fig.~\ref{fig1}(b)]. Neurons were identified by their location within the midbrain slice preparation and for a subset of recorded neurons we obtained additional identification via intracellular fills \cite{Wang04,Wang06}.

\begin{figure*}[t]
  \begin{center}
    \epsfig{file=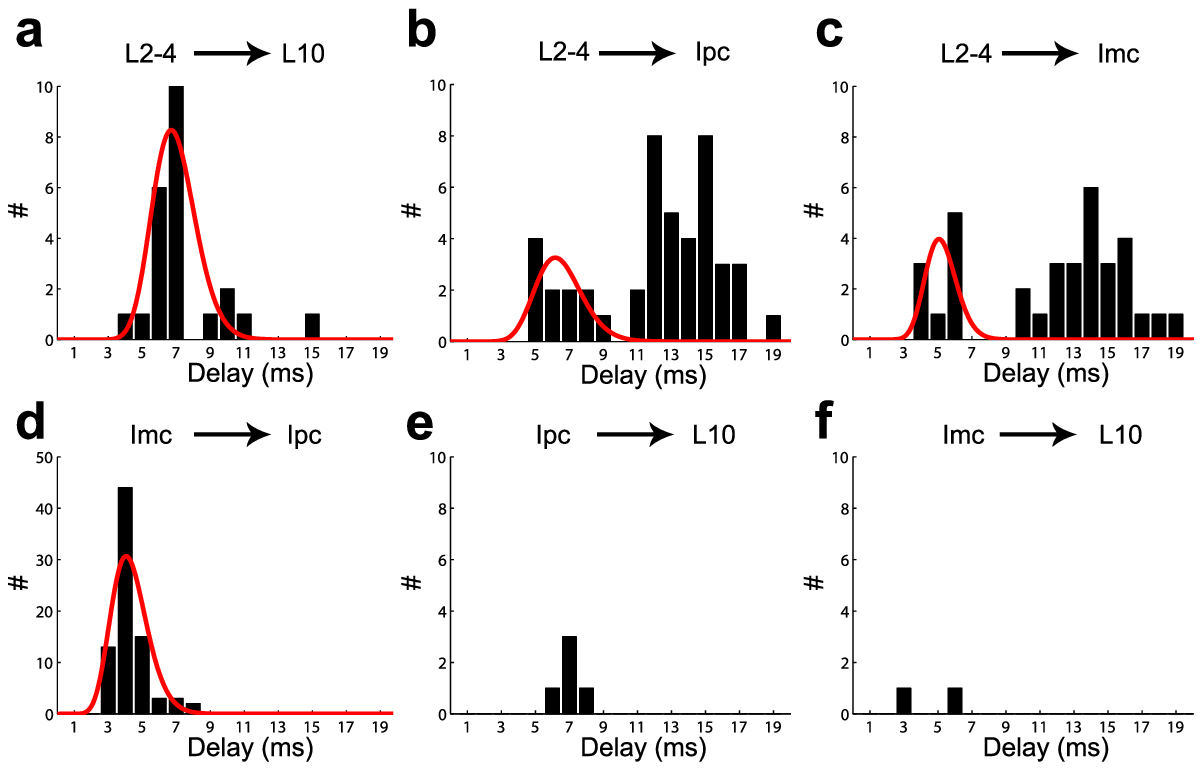,width=4in}    
    \caption{(Color) Measured distribution of signal delays between isthmotectal elements and plot of the corresponding gamma distribution [red curves in {\bf (a)} through {\bf (d)}] with the same mean and standard deviation. {\bf (a)} L2-4 to L10. {\bf (b)} L2-4 to Ipc. {\bf (c)} L2-4 to Imc. {\bf (d)} Imc to Ipc. {\bf (e)} Ipc to L10. {\bf (f)} Imc to L10.}
    \label{fig2}
  \end{center}
\end{figure*}

A subpopulation of tectal layer 10 (L10) neurons project to both the ipsilateral Ipc and Imc in a topographic fashion \cite{Wang04,Wang06,Cajal,Hunt76,Hunt77,Woodson}. Their apical dendrite courses straight up to layer 2 with few ramifications, and basal dendrites reach down to the border of layer 13. Retinal axon terminals overlap with the apical dendrite in tectal layers 2 to 7 \cite{Domesick,Sebesteny}. We placed a stimulus electrode in layer 2 to 4 (L2-4) and recorded from L10 neurons with whole-cell recordings in response to L2-4 stimulation. The delays from the beginning of the stimulus pulse to the onset of the L10 response ranged from 4 to 15 ms with a mean delay of 6.9 ms and a standard deviation (SD) of 1.3 ms ($n = 15$ cells) [Fig.~\ref{fig2}(a)]. Tectal L10 neurons are a heterogeneous population \cite{Wang06}. Therefore, only filled L10 neurons with axons originating from the dendrite were included in this analysis. Since L10 neuron dendrites can reach up to L2, the possibility of unwanted direct electrical, rather than synaptic, stimulation of L10 neuron dendritic endings arises. At the end of a recording session, we evaluated the nature of stimulation by blocking chemical synaptic transmission via the block of Ca-channels by replacing Ca$^{2+}$ in the saline with Mg$^{2+}$. 

We measured signal delays between optic tectum and individual Ipc neurons via RGC axon stimulation or L10 neuron dendrite stimulation, with a stimulus electrode placed in tectal L2-4. In the first case, the group of stimulated RGC axons stimulates a population of L10 neurons, which in turn stimulates a large number of Ipc neurons. In the second case, L10 neurons are stimulated directly. This stimulus paradigm provided a high chance of recording from an Ipc neuron that received tectal synaptic inputs. The delays from the beginning of the stimulus pulse to the onset of the Ipc neuron response ranged from 5 to 19 ms ($n = 17$ cells) [Fig.~\ref{fig2}(b)]. As expected from the stimulus paradigm, the distribution of delays is bimodal. We suspect that the first bump (5 to 9 ms range) is dominated by direct L10 dendrite stimulation (mono-synaptic pathway L10-Ipc); whereas the second bump (11 to 19 ms range) is dominated by RGC axon stimulation, which initiates the bi-synaptic pathway RGC-L10-Ipc. From the first bump in the histogram we estimate a mean delay of 6.5 ms and a SD of 1.4 ms for the mono-synaptic pathway L10-Ipc. Since Ipc neuron axons can reach up to L2 \cite{Wang06}, the possibility of unwanted direct electrical stimulation of Ipc axons arises. At the end of a recording session, we evaluated the nature of stimulation by blocking chemical synaptic transmission via replacing Ca$^{2+}$ in the saline with Mg$^{2+}$. 

Using a stimulus paradigm similar to the one described above, we measured signal delays between L10 and individual Imc neurons. We placed a stimulus electrode in L2-4 for stimulation of RGC axons or L10 neuron dendrites and recorded from Imc neurons with whole-cell recordings in response to L2-4 stimulation. The signal delays ranged from 4 to 19 ms ($n = 17$ cells) and the distribution was bimodal [Fig.~\ref{fig2}(c)]. As described above, the first bump is likely to be dominated by the mono-synaptic pathway (L10-Imc), whereas the second bump is likely to be dominated by the bi-synaptic pathway (RGC-L10-Imc). The first bump in the histogram yielded a mean delay of 5.2 ms and a SD of 0.9 ms. Since Imc axons terminate in tectal layers 10 to 13 \cite{Wang04}, the possibility of direct Imc axon stimulation via stimulus electrodes in L2-4 does not arise.

The Imc nucleus consists of two cell types, one of which projects to the Ipc nucleus with a broad and dense projection of axonal arbors \cite{Wang04,Tombol95,Tombol98}. We positioned a stimulus electrode in the Imc nucleus and recorded from Ipc neurons with whole-cell recordings in response to Imc stimulation. The signal delays ranged from 3 to 8 ms with a mean delay of 4.3 ms and a SD of 1.1 ms ($n = 12$ cells) [Fig.~\ref{fig2}(d)]. Care had to be taken about the interpretation of the Imc stimulation experiments. The stimulus electrode in the Imc nucleus stimulates 4 elements: L10 neuron axons, Ipc neuron axons passing through the Imc nucleus, and two populations of Imc neurons; one projecting to tectum and the other projecting to Ipc. To filter out the Imc to Ipc synaptic connection, we stimulated in an area of the Imc nucleus that did not correspond to the topographic location of the recorded Ipc neuron, thus avoiding both antidromic stimulation of the axon from the recorded Ipc neuron as well as avoiding orthodromic stimulation of the L10 axons passing through the Imc nucleus on their way to the same location in the Ipc nucleus. At the end of a recording session, we applied bicuculline to verify that the synaptic inputs to the recorded Ipc neuron were indeed from the stimulated GABAergic Imc neurons (GABA: gamma-aminobutyric acid). The responses disappeared when 100 $\mu$M bicuculline was added to the bath (data not shown) thus (i) indicating that the responses were of synaptic origin (rather than antidromic Ipc or L10 axon stimulation) and (ii) confirming that GABA is the transmitter as had been suggested by anatomical studies \cite{Wang04}.

The Ipc nucleus has topographical reciprocal connections with the tectum \cite{Wang06,Hunt76,Hunt77,Gunturkun,Hellmann}. The efferents from Ipc have large calibre axons and terminate in a columnar manner ranging from layers 2 to 12 [Fig.~\ref{fig1}(a)] \cite{Wang06,Cajal,Hunt76,Hunt77,Tombol95,Tombol98}. We applied local extracellular electrical stimulation of a group of Ipc neurons with a stimulus electrode placed in the Ipc nucleus. Such extracellular electrical stimulation also stimulates L10 axons antidromically. The fast L10 neuron antidromic responses were distinguishable from the much slower and long-lasting synaptic responses. The additional direct activation of Imc axons in the Ipc nucleus does not interfere with this experiment, since the population of Imc neurons projecting to the Ipc nucleus is different from the population of Imc neurons projecting to the tectum. The yield for finding Ipc to L10 synaptic responses turned out to be very low. For the few cases we found, the delays ranged from 6 to 8 ms ($n = 5$ cells) [Fig.~\ref{fig2}(e)].

The projection from individual Imc neurons to tectal layers 10 to 13 is broad and sparse \cite{Wang04}. We positioned a stimulus electrode in the Imc nucleus and recorded from L10 neurons with whole-cell recordings in response to Imc stimulation. The yield for finding Imc to L10 synaptic responses turned out to be very low. For the two connected pairs we found, the signal delays were 3 and 6 ms ($n = 2$ cells) [Fig.~\ref{fig2}(f)]. The low yield and the interpretation of these experiments require some explanation. As mentioned above, a stimulus electrode in the Imc nucleus will stimulate four elements. To filter out the Imc to L10 synaptic connection, we stimulated in an area of the Imc nucleus that did not correspond to the topographic location of the recorded L10 neuron, thus avoiding both antidromic stimulation of the axon from the recorded L10 neuron as well as avoiding orthodromic stimulation of the Ipc axons passing through the Imc nucleus on their way to the same location of the tectum. At the end of a recording session, we applied bicuculline to verify that the synaptic inputs to the recorded L10 neuron were indeed from the stimulated GABAergic Imc neurons. For the two neurons, the responses disappeared when 100 $\mu$M bicuculline was added to the bath (data not shown) thus indicating that the responses were of synaptic origin; rather than antidromic L10 or orthodromic Ipc axon stimulation.

In summary, these data show that the signal delays between isthmotectal elements are distributed ranging from 4 to 9 ms.
\begin{figure}[t]
  \begin{center}
    \epsfig{file=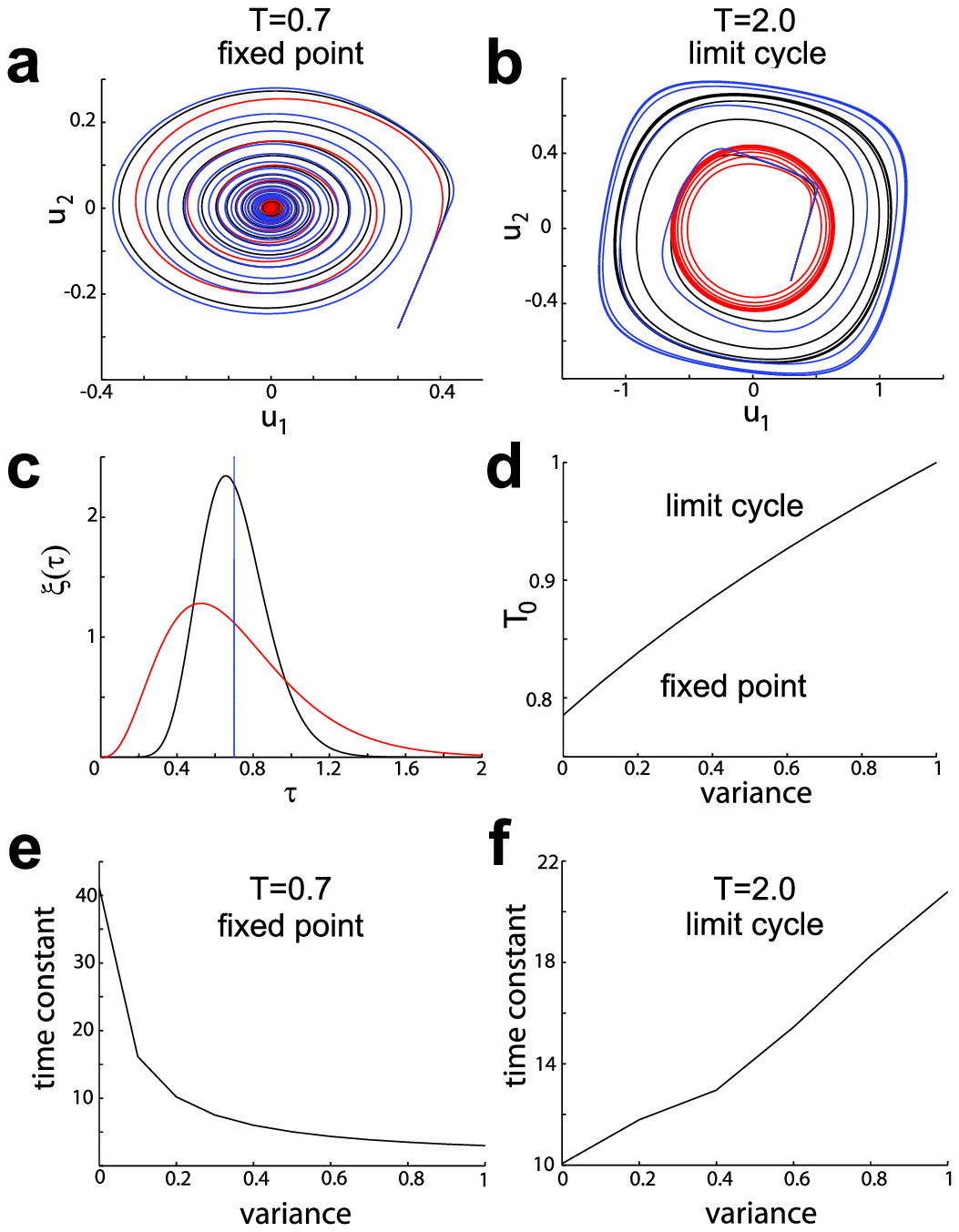,width=3in}    
    \caption{(Color) Mean delays and attractors. {\bf (a)}, {\bf (b)} Dynamics of the two-neuron model system for gamma distributions with mean delay values of $T = 0.7$ [{\bf (a)}, fixed point] and   [{\bf (b)}, limit cycle], respectively. For both cases, the standard deviation is 0\% (blue), 25\% (black), and 50\% (red) of the mean delay. The initial condition is $u_1(t) = 0.30$ and $u_2(t) = -0.28$  for $-\tau \leq t \leq 0$. {\bf (c)} Gamma distribution for a mean delay value of $T = 0.7$ and a standard deviation of 0\% (blue), 25\% (black), and 50\% (red) of the mean delay. {\bf (d)} Critical mean delay, $T_0$, where the Hopf bifurcation takes place, plotted against variance. {\bf (e)}, {\bf (f)} Time constant for reaching the attractor for $T = 0.7$ (fixed point) and $T = 2.0$ (limit cycle), respectively, plotted against the variance of the gamma distribution.}
    \label{fig3}
  \end{center}
\end{figure}   

\section{Distributed delays and the dynamics of neural feedback systems}
What is the impact of distributed delays on a mathematically tractable neural model feedback system? To interpret the potential impact of the measured distribution of delays on the dynamics of neural feedback systems, we investigated a model system of two coupled Hopfield neurons \cite{Babcock,Brandt06b,Brandt07a,Hopfield}, described by the first-order delay differential equations
\begin{eqnarray}
\frac{du_1(t)}{dt} &=& - u_1(t)+a_1 \tanh[u_2(t-\tau_2)]\, , \nonumber \\  
\frac{du_2(t)}{dt} &=& -u_2(t)+a_2 \tanh[u_1(t-\tau_1)] \, . \label{eq12}
\end{eqnarray}
Here, $u_1(t)$ and $u_2(t)$ denote the voltages of the model neurons and $\tau_1$ and $\tau_2$ are the temporal delays, while $a_1$ and $a_2$ describe the coupling strength between the two neurons. In this analysis, the time variable is dimensionless. Translation to real time can be made by multiplying the dimensionless time variable with a membrane time constant, $RC$. The system of delay differential equations has a trivial stationary point at the origin, $u_1=u_2=0$ [Fig.~\ref{fig3}(a)]. For $a_1 a_2 \leq -1$, the fixed point at the origin is asymptotically stable as long as the mean of the time delays $(\tau_1 + \tau_2)/2$ does not exceed a critical value $\tau_0$ \cite{Brandt06b,Wei}:  
\begin{eqnarray}
\frac{\tau_1 + \tau_2}{2} < \tau_0 = \frac{1}{2 \sqrt{|a_1 a_2| - 1}} \sin^{-1}\frac{2\sqrt{|a_1 a_2| - 1}}{|a_1 a_2|} \label{eq3}\, .
\end{eqnarray}
The critical value $\tau_0$ is determined by combinations of the product of the couplings alone [Eq.\ (\ref{eq3})]. For couplings of opposite signs (e.g., $a_1a_2 \leq -1$) and when the delays are increased, the origin becomes unstable and a limit cycle emerges via a supercritical Hopf bifurcation at $(\tau_1 + \tau_2)/2 = \tau_0$ [Fig.~\ref{fig3}(b)]. The critical value, $\tau_0$ , decreases with decreasing value of the product of the couplings $a_1a_2$ below $-1$. In other words, oscillations can be achieved by either increasing the delays or by increasing the absolute value of the coupling strengths of opposite signs.

For a distribution of delays we replace the coupling term in (\ref{eq12}) with a weighted sum over similar terms but with different delays
\begin{eqnarray}
\frac{du_1(t)}{dt} &=& - u_1(t) + a_1 \int_0^{\infty} d \tau \xi(\tau) \tanh[u_2(t - \tau)]\, , \nonumber \\
\frac{du_2(t)}{dt} &=& - u_2(t) + a_2 \int_0^{\infty} d \tau \xi(\tau) \tanh[u_1(t - \tau)]\, . \label{eq45}
\end{eqnarray}
The delay kernel $\xi(\tau)$ is normalized to satisfy $\int_0^{\infty}d\tau \xi(\tau) = 1$ . For simplicity, we chose the delay kernels to be identical for both legs of the loop. We chose the delay kernel to be a gamma distribution,
\begin{eqnarray}
\xi(\tau) = \frac{(T / \nu )^{T^2 / \nu}}{\Gamma(T^2 / \nu)} \tau^{T^2/ \nu - 1}e^{- T \tau / \nu} \, ,
\end{eqnarray}
where $T$ is the mean delay, $\nu$ is the variance of the gamma distribution, and the gamma function is defined as $\Gamma(x) = \int_0^{\infty}t^{x-1}e^{-t}dt$. The gamma distribution was chosen because it has the biologically plausible feature to vanish for delays approaching $0$ [Fig.~\ref{fig3}(c)]. For the coupling strengths we chose $a_1 = -2$  and $a_2 = 1$ for all simulations.
\begin{figure}[t]
  \begin{center}
    \epsfig{file=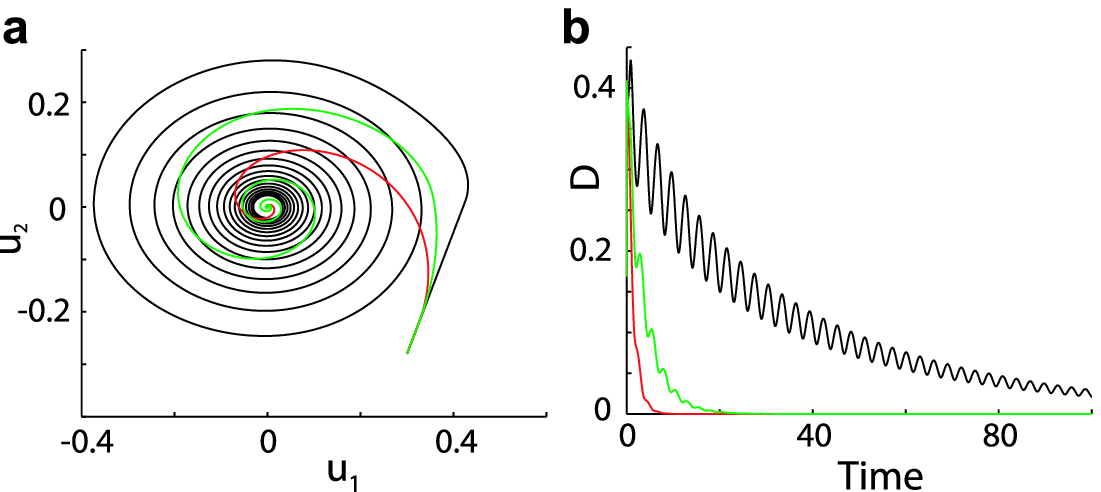,width=3.4in}    
    \caption{(Color) Dynamics of the two-neuron model system with discrete delays. {\bf (a)} Dynamics of the system with two fixed delays of 0.1 and 0.7 (green); one fixed delay of 0.1 (red); and one fixed delay of 0.7 (black). {\bf (b)} Distance $D(t) = \sqrt{u_1^2(t) + u_2^2(t)}$ from (0, 0) vs.\ time for the above cases.}
    \label{fig4}
  \end{center}
\end{figure}   

The parameters to vary are the mean delay, $T$, and the variance,  $\nu$, of the gamma distribution. As these parameters are changed, the fixed point at the origin changes from a stable fixed point to an unstable fixed point surrounded by a stable limit cycle and vice-versa (Hopf bifurcation). This transition takes place when the roots, $\lambda$, of the characteristic equation for the system (\ref{eq45}),
\begin{eqnarray}
\lambda^2 + 2 \lambda + 1 - a_1 a_2 \left(1 + \frac{\lambda \nu}{T}\right)^{2T^2 / \nu} = 0\, ,
\end{eqnarray}
are purely imaginary. The characteristic equation is obtained by demanding that the solution to (\ref{eq45}) behaves as $u_1(t) = Ae^{\lambda t}$, $u_2(t) = B e ^{\lambda t}$ near the fixed point. Substituting $\lambda = i \omega$, where $\omega$ is real, we have
\begin{eqnarray}
-\omega^2 + 2 i \omega + 1 - a_1 a_2\left( 1 + \frac{i \omega \nu}{T}\right)^{-2T^2/\nu} = 0 \, .
\end{eqnarray}
Separating real and imaginary parts, we get two equations from which we can numerically eliminate $\omega$. However, there are multiple solutions for this. For a given variance $\nu$, the solution with the minimum positive mean delay $T$, determines the critical mean delay $T_0$ at which the fixed point at the origin loses its stability and a stable limit cycle emerges. Our analysis shows that the introduction of distributed delays (increasing variance) leads to a smaller limit cycle [Fig.~\ref{fig3}(b)]. Furthermore, the critical mean delay $T_0$ increases with increasing variance [Fig.~\ref{fig3}(d)].

To estimate the time constant for reaching an attractor, we calculated the distance $D_{\theta}(t) = \sqrt{u_1^2(t) + u_2^2(t)}$ from the origin along a given polar angle $\theta$ in the $u_1$-$u_2$ space. Assuming an exponential dependence, a fit of an exponential function to the simulated $D_{\theta}(t)$ values provided the time constant for that polar angle. We repeated the procedure for 360 polar angles in 1-deg increments and took the final time constant to be the mean of the 360 time constants at given polar angles. This analysis shows that increasing variance makes the convergence to the fixed points faster [Fig.~\ref{fig3}(e)] and the convergence to limit cycles slower [Fig.~\ref{fig3}(f)]. 

In summary, distributed delays increase the parameter region with fixed-point behavior and accelerate the convergence to the fixed point.

The enhanced fixed-point stability with distributed delays actually comes from the contribution of the smaller delays, rather than the distribution per se. To illustrate this insight, we compare the dynamics of three systems with delays given by one or two superimposed delta-distributions with peaks at: (i) $\tau = 0.7$, (ii) both $\tau=0.1$  and $\tau = 0.7$, and (iii) $\tau = 0.1$ (Fig.~\ref{fig4}). The distributed system (ii) converges faster than system (i), but the distributed system (ii) converges slower than system (iii). In other words, adding a longer delay of $\tau=0.7$ to the $\tau=0.1$ system slows the convergence, whereas adding a shorter delay of $\tau=0.1$ to the $\tau=0.7$ system accelerates the convergence. Thus, it is not the distribution of delays per se, but the contribution of shorter delays in the distribution that enhances fixed-point stability.

\section{Discussion}
For large brains with finite signal propagation velocities, delays are a fact of life. In some feedforward pathways, such as the vertebrate optic nerve, delays can be specific to the retinal ganglion cell type thus leading to differences of arrival time for different retinal representations of the visual stimulus \cite{Hoffman,Mpodozis,Northmore,Letelier}. In other feedforward sensory pathways, such as the avian nucleus laminaris, delays are used explicitly to evaluate interaural time differences \cite{Carr}. Delays in feedback loops play a fundamentally different role, as they can determine the dynamical behavior of the system \cite{Milton,Fisher}. Specifically, for delays smaller than a critical value a neural feedback system may converge toward a steady-state, whereas for delays larger than the critical value the system may oscillate \cite{Coleman,Heiden}. In nonlinear systems, the distribution of a system parameter can have unexpected effects on the systems dynamics \cite{Braiman,Brandt06a,Chacon}. Consequently, if delay is a relevant parameter in neural feedback systems, as stated above, it is important to investigate the impact of delay distributions on the system dynamics. 

Parameters in biological system are usually distributed over some range. Therefore, these systems must be robust in the sense that the parameter variability should not detain the biological system from functioning correctly. Moreover, a system architecture in which the parameter variability actually enhances the system's performance would be particularly desirable. In this study, we have quantified the distribution of delays in the avian isthmotectal feedback loop. Furthermore, by investigating a mathematical model of coupled neurons with distributed delays, we have demonstrated that distributed delays enhance the stability of the system, where the stabilizing effect arises from the contribution of smaller delays introduced through the delay distribution. Since the functional role of the isthmotectal feedback loop remains mostly unclear to date, it is not obvious whether this stabilizing effect is beneficial to the system's functioning. Further understanding of the neuronal processes in the isthmotectal feedback loop will be necessary to answer this question.

\section{Discussion}
For large brains with finite signal propagation velocities, delays are a fact of life. In some feedforward pathways, such as the vertebrate optic nerve, delays can be specific to the retinal ganglion cell type thus leading to differences of arrival time for different retinal representations of the visual stimulus \cite{Hoffman,Mpodozis,Northmore,Letelier}. In other feedforward sensory pathways, such as the avian nucleus laminaris, delays are used explicitly to evaluate interaural time differences \cite{Carr}. Delays in feedback loops play a fundamentally different role, as they can determine the dynamical behavior of the system \cite{Milton,Fisher}. Specifically, for delays smaller than a critical value a neural feedback system may converge toward a steady-state, whereas for delays larger than the critical value the system may oscillate \cite{Coleman,Heiden}. In nonlinear systems, the distribution of a system parameter can have unexpected effects on the systems dynamics \cite{Braiman,Brandt06a,Chacon}. Consequently, if delay is a relevant parameter in neural feedback systems, as stated above, it is important to investigate the impact of delay distributions on the system dynamics. 

Parameters in biological system are usually distributed over some range. Therefore, these systems must be robust in the sense that the parameter variability should not detain the biological system from functioning correctly. Moreover, a system architecture in which the parameter variability actually enhances the system's performance would be particularly desirable. In this study, we have quantified the distribution of delays in the avian isthmotectal feedback loop. Furthermore, by investigating a mathematical model of coupled neurons with distributed delays, we have demonstrated that distributed delays enhance the stability of the system, where the stabilizing effect arises from the contribution of smaller delays introduced through the delay distribution. Since the functional role of the isthmotectal feedback loop remains mostly unclear to date, it is not obvious whether this stabilizing effect is beneficial to the system's functioning. Further understanding of the neuronal processes in the isthmotectal feedback loop will be necessary to answer this question.

\section{Experimental methods}
White Leghorn chick hatchlings (Gallus gallus) of less than 3 days of age were used in this study. All procedures used in this study were approved by the local authorities and conform to the guidelines of the National Institutes of Health on the Care and Use of Laboratory Animals. Animals were injected with ketamine (40 mg per kg, i.m.). Brain slices of the midbrain were prepared following published protocols \cite{Dye,Luksch98,Luksch01,Luksch04,Khanbabaie}. Briefly, preparations were done in $0$ \textcelsius, oxygenated, and sucrose-substituted saline (240mM sucrose, 3 mM KCl, 5 mM MgCl$_2$, 0.5 mM CaCl$_2$, 1.2 mM NaH$_2$PO$_4$, 23 mM NaHCO$_3$, and 11 mM D-glucose). After decapitation, the brains were removed from the skull, and the forebrain, cerebellum, and medulla oblongata were discarded. A midsagittal cut was used to separate the tectal hemispheres. The tectal hemispheres were sectioned at 500 $\mu$m on a tissue slicer (Vibroslice, Camden and VF-200, Precisionary Instruments) in either the transverse or the horizontal plane. Slices were collected in oxygenated saline (120 mM NaCl, 3 mM KCl, 1 mM MgCl$_2$, 2 mM CaCl$_2$, 1.2 mM NaH$_2$PO$_4$, 23 mM NaHCO$_3$, and 11 mM D-glucose) and kept submerged in a chamber that was bubbled continuously with carbogen (95\% oxygen, 5\% CO$_2$) at room temperature. The slice was then transferred to a recording chamber (RC-26G, Warner Instruments) mounted on a fixed stage upright microscope equipped with DIC optics (BX-51WI, Olympus). The slice was held gently to the bottom of the chamber with an anchor of nylon threads, and the chamber was perfused continuously with oxygenated saline at room temperature. The cells in L10, Imc, and Ipc are visible with DIC optics.

Local electrostimulation was achieved by inserting bipolar tungsten electrodes under visual control into either the upper tectal retinorecipient layers (2 to 4), layer 5b, or the isthmic nuclei Ipc or Imc with a three-axis micromanipulator (U-31CF, Narishige). Electrodes were custom-built from 50-$\mu$m diameter, insulated tungsten wires (California Fine Wire) that were glued together with cyanoacrylate and mounted in glass microcapillaries for stabilization. The wires protruded several hundred µm from the capillaries, and the tips were cut at an angle. Stimulus isolators (Isolated Pulse Stimulator 2100, AM Systems) generated biphasic current pulses (20 - 200 $\mu$A, 500 $\mu$s).

Whole-cell recordings were obtained with glass micropipettes pulled from borosilicate glass (1.5 mm OD, 0.86 mm ID, AM Systems) on a horizontal puller (P-97, Sutter Instruments and DMZ Universal Puller, Zeitz Instruments) and were filled with a solution containing 100 mM K-Gluconate, 40 mM KCl, 10 mM HEPES, 0.1 mM CaCl$_2$, 2 mM MgCl$_2$, 1.1 mM EGTA, 2 mM Mg-ATP, pH adjusted to 7.2 with KOH. Electrodes were advanced through the tissue under visual guidance with a motorized micromanipulator (MP-285, Sutter Instruments) while constant positive pressure was applied and the electrode resistance was monitored by short current pulses. Once the electrode had attached to a membrane and formed a seal, access to the cytosol was achieved by brief suction. Whole-cell recordings were performed with the amplifier (Axoclamp 2B, Axon Instruments and SEC-05L, npi-electronic) in the bridge mode (current clamp). The series resistance was estimated by toggling between the bridge and the DCC (discontinuous current clamp) mode. The series resistance was compensated with the bridge balance. Analog data were low-pass filtered (4-pole Butterworth) at 1 kHz, digitized at 5 kHz, stored, and analyzed on a PC equipped with an PCI-MIO-16E-4 and LabView software (both National Instruments).

Labeling of a subset of recorded neurons was carried out as described previously \cite{Luksch98,Luksch01,Luksch04,Mahani}. In brief, whole-cell patch recordings were obtained as described above. Additionally, the electrode solution contained 0.5\% Biocytin (w/v) to label the recorded neurons. Individual cells were filled intracellularly with 2 nA of positive current over 3 minutes. After recording and labeling, slices were kept in oxygenated ACSF for an additional 30 minutes and subsequently fixed by immersion in 4\% paraformaldehyde in PB for at least 4 hours. Slices were then washed in phosphate buffer (PB, 0.1 M, pH 7.4) for at least 4 hours, immersed in 15\% sucrose in PB for at least 4 hours and then immersed in 30\% sucrose in PB for 12 hours, and resectioned at 60 $\mu$m on a freezing microtome. The sections were collected in PB and the endogenous peroxidase blocked by a 15-minute immersion in 0.6\% hydrogen peroxide in methanol. The tissue was washed several times in PB, and then incubated in the avidin-biotin complex solution (ABC {\it Elite} kit, Vector Labs) and the reaction product visualized with a heavy-metal intensified DAB protocol. Following several washes in PB, the 60 $\mu$m-thick sections were mounted on gelatin-coated slides, dried, dehydrated, and coverslipped. Sections were inspected for labeled neurons, and only data from cells that could unequivocally be classified according to published criteria \cite{Wang04,Wang06} were taken for further analysis. Cells were reconstructed at medium magnification (10x to 20x) with a camera lucida on a Leica microscope and projected onto the 2D plane.

\section{Acknowledgements}
The authors thank Edward Gruberg for critical reading of the manuscript. The work was supported by grant Lu 622 8-2 DFG to H.\ L.\ and by grant NIH R01 EY15678 to R.\ W.


%
\end{document}